\documentclass[aps,prl,twocolumn,prl,superscriptaddress]{revtex4-2}
\usepackage[dvipdfmx]{graphicx}
\usepackage{bm}
\usepackage{color}
\usepackage{ulem}

\begin{document}


\title{Periodic super-radiance in Er:YSO crystal}


\author{Hideaki~Hara}
\email{hhara@okayama-u.ac.jp}
\affiliation{Research Institute for Interdisciplinary Science, Okayama University, Okayama 700-8530, Japan}
\author{Junseok~Han}
\affiliation{Research Institute for Interdisciplinary Science, Okayama University, Okayama 700-8530, Japan}
\affiliation{Department of Physics and Astronomy \& Institute of Applied Physics, Seoul National University, Seoul 08826, Korea}
\author{Yasutaka~Imai}
\affiliation{Research Institute for Interdisciplinary Science, Okayama University, Okayama 700-8530, Japan}
\author{Noboru~Sasao}
\affiliation{Research Institute for Interdisciplinary Science, Okayama University, Okayama 700-8530, Japan}
\author{Akihiro~Yoshimi}
\affiliation{Research Institute for Interdisciplinary Science, Okayama University, Okayama 700-8530, Japan}
\author{Koji~Yoshimura}
\affiliation{Research Institute for Interdisciplinary Science, Okayama University, Okayama 700-8530, Japan}
\author{Motohiko~Yoshimura}
\affiliation{Research Institute for Interdisciplinary Science, Okayama University, Okayama 700-8530, Japan}
\author{Yuki~Miyamoto}
\email{miyamo-y@cc.okayama-u.ac.jp}
\affiliation{Research Institute for Interdisciplinary Science, Okayama University, Okayama 700-8530, Japan}



\date{\today}

\begin{abstract}
We observed periodic optical pulses from an Er:YSO crystal 
during irradiating with an continuous-wave excitation laser.
We refer to this new phenomenon as "periodic super-radiance".
This periodicity can be understood qualitatively 
by a simple model, in which 
a cyclic process of a continuous supply of population inversion and a sudden burst of super-radiance is repeated.
The excitation power dependences of 
peak interval and the pulse area 
can be interpreted with our simple model.
In addition, the linewidth of super-radiance 
is much narrower than an inhomogeneous broadening in a crystal.
This result suggests that only Er$^{3+}$ ions in a specific environment 
are involved in super-radiance.
\end{abstract}


\maketitle


A considerable number of theoretical and experimental studies on super-radiance (SR) have been conducted, 
starting with the prediction by R. H. Dicke in 1954 \cite{SR-1954}.
If a correlation between the atomic or molecular dipole moments is generated, 
the peak transition rate is accelerated and gets proportional to $N^{2}$ 
because a macroscopic dipole proportional to the number of atoms $N$ is created.
For SR to occur, 
this macroscopic dipole has to develop within a decoherence time $T_{2}$ 
and a population inversion has to reach a certain threshold.
So far 
SR has been demonstrated in gases \cite{SR-HFgas-1973,cascadeSR-Na-1976,triggeredSR-1980} 
and solid-state materials \cite{SR-O2KCl-1982,SR-nanocrystal-2018,SR-NVcenter-2018,SR-solid-review-2016}.

Recently, SR was observed from Er$^{3+}$ ions 
doped in a Y$_{2}$SiO$_{5}$ crystal (Er:YSO) \cite{ErYSO-Padova-2020} 
\footnote{In some literatures, as in  \cite{ErYSO-Padova-2020}, 
SR is referred to as super-fluorescence when it starts from uncorrelated fully excited states.}.
The intraconfiguration $4f \to 4f$ transitions of rare-earth ions doped into crystals 
has remarkably long decoherence time 
despite being in a solid environment 
because $4f$ electrons are shielded by $5s$ and $5p$ shells from interactions with the host lattice.
An Er:YSO crystal exhibits a narrowest homogeneous linewidth of 73 Hz 
in a solid \cite{ErYSO-linewidth-2009}.
Various applications towards quantum information processing 
\cite{quantum-memory-2010,RE-quantum-memory-2015,RE-quantum-memory-2018} 
and fundamental physics \cite{axion-2018,RANP-2019,MagRENP-2021} 
have been proposed.

In this paper, we report on experimental studies of SR 
with a quasi-periodic time structure observed in an Er:YSO crystal.
In past SR experiments, 
usually a single SR pulse, 
with or without ringing, 
is generated after an excitation by a pulsed laser.
In contrast, observed pulses in the present case are generated 
periodically 
while the continuous-wave (CW) excitation laser is turned on.
In addition, its linewidth is found to be comparable to the Fourier limit, 
much narrower than an inhomogeneous broadening in a crystal.
This phenomenon is not only interesting in itself, but also 
from the viewpoint that optical pulses with a narrow linewidth are generated for an input of a CW excitation laser.
To our best knowledge, such phenomena have not been studied experimentally or theoretically up to now.

Below we describe the results of our experiment, focusing on the periodic nature. 
A simple model is also given which treats it as 
repetition of a cyclic process of a continuous supply of population inversion 
and a sudden burst of SR emissions after reaching an SR threshold.

\begin{figure}
\centering
\includegraphics[width=8.5cm,keepaspectratio]{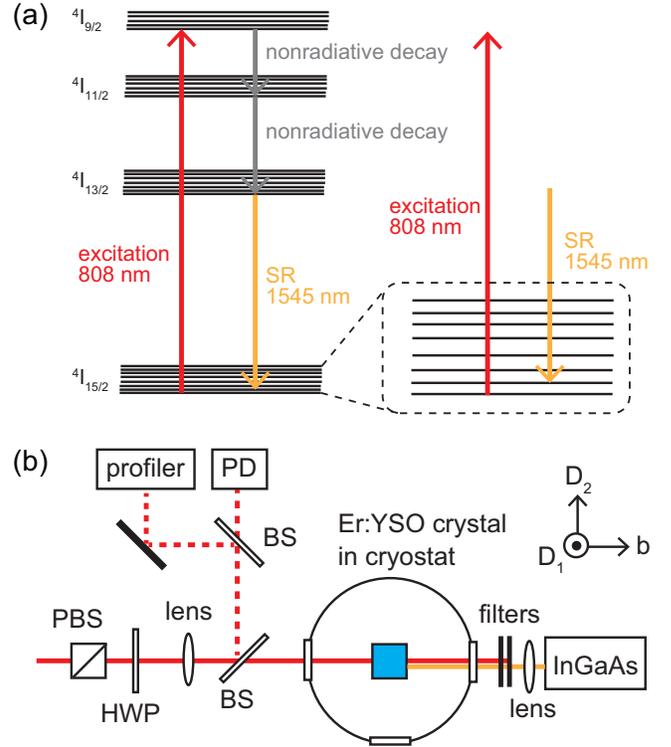}
\caption{(a) Energy diagram of Er$^{3+}$ ion doped in YSO crystal. 
The dashed rounded square shows an enlarged view of the $^4$I$_{15/2}$ ground state.
(b) Experimental setup. 
PBS ; polarization beam splitter. 
HWP ; half-wave plate.
BS ; beam splitter.
PD ; photodetector.
InGaAs ; InGaAs photodetector.}
\label{fig:setup}
\end{figure}

Figure \ref{fig:setup} (a) shows the energy diagram of Er$^{3+}$ ions doped in a YSO crystal.
Each state is split into 
several Stark levels by the crystal field.
Er$^{3+}$ ions are excited from the ground state to the lowest Stark level of the $^4$I$_{9/2}$ state 
with a wavelength of 808 nm.
The excited Er$^{3+}$ ions rapidly decay via nonradiative processes.
In this way the population accumulates in the lowest Stark level of the $^4$I$_{13/2}$ state with a long lifetime, 
which is measured by the fluorescence (spontaneous emission) to be roughly 10 ms.
Besides the fluorescence, the optical pulses propagating in the forward direction are generated.

Figure \ref{fig:setup} (b) shows the experimental setup.
The target is an Er$^{3+}$:YSO crystal grown by Scientific Materials, Inc. of Bozeman, Montana.
The Er$^{3+}$ ions at site 2 are used in our experiment.
The number density of Er$^{3+}$ ions at each site is roughly $5\times10^{18}$ cm$^{-3}$ 
according to the nominal concentration value of 0.1 \%.
The crystal is held by a copper holder in a cryostat 
and it is cooled to roughly 4 K by a Gifford-McMahon refrigerator.
Almost all the Er$^{3+}$ ions exist in the lowest Stark level of the $^4$I$_{15/2}$ ground state at 4 K.
The crystal is aligned with the laser $\mathbf{k}$ vector parallel to the $\mathbf{b}$ axis 
along a 6 mm path. 
Other crystal dimensions are 
4 mm along $\mathbf{D}_{1}$ axis and 5 mm along $\mathbf{D}_{2}$ axis.
The polarization of the input laser is parallel to $\mathbf{D}_{2}$ axis.
The beam diameter ($2 w_{0}$) is loosely focused on the crystal to roughly 300 $\mu$m 
and it remains almost unchanged in the crystal. 
The input laser power is monitored by a photodetector and it is typically 90 mW.
The laser position is also monitored by a beam profiler.
An indium-gallium-arsenide (InGaAs) photodetector (HCA-S-200M-IN; FEMTO Messtechnik GmbH) 
detects generated optical pulses.
The residual excitation laser is removed by long pass filters in front of the detector.

The time sequence of the laser excitation of Er$^{3+}$ ions is controlled by an acousto-optic modulator (AOM).
The excitation laser is turned on for 40 ms every 200 ms: 
 this time sequence is chosen in such a way that 
 the duration of the excitation laser is much longer than the observed pulse duration and period, 
 and at the same time the temperature rise of the target crystal, 
which shortens the decoherence time, 
is enough low.  
The 40 ms excitation data is taken 10 times repeatedly.


\begin{figure}
\centering
\includegraphics[width=8.5cm,keepaspectratio]{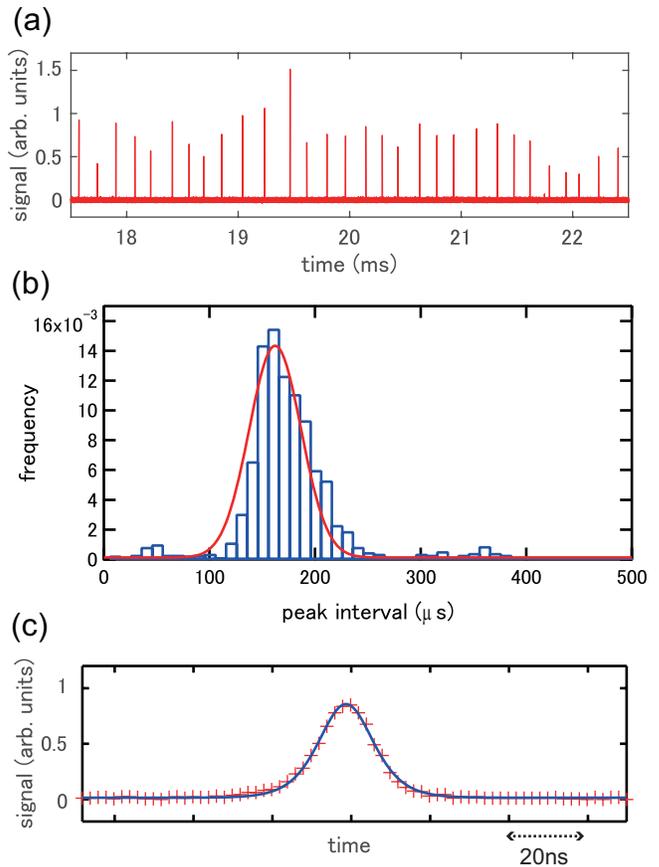}
\caption{(a) Example of waveform of observed periodic pulses. 
The excitation laser is turned on for 40 ms from $t=0$.
(b) Histogram of the peak interval between neighboring pulses.
The red line is the Gaussian fit.
(c) Example of observed single pulse. 
The red crosses are the data points.
The blue solid line is the fit by a sech-squared function.}
\label{fig:SR}
\end{figure}

 We observed a train of light pulses: 
 the pulses begin to appear several ms 
after the laser is turned on ($t=0$), and disappears 
 right after it is turned off.
Figure \ref{fig:SR} (a) shows an example of the observed pulses in the middle 5 ms 
of the excitation.
The pulses are generated at almost constant time intervals, in other words, periodically.
Figure \ref{fig:SR} (b) is a histogram of peak intervals, 
which are 
 defined by the time difference between adjacent peaks 
 (the data after $t=10$ ms are plotted, 
  where the system has reached equilibrium).
The red line is the Gaussian fit.
The mean period and the standard deviation given by the fit are 160 $\mu$s and 20 $\mu$s, respectively.
The pulses are generated in the transition from the lowest Stark level of the $^{4}$I$_{13/2}$ state 
to the second lowest Stark level of the $^{4}$I$_{15/2}$ ground state.
It is confirmed by a monochromator and the wavelength of the pulses is $\lambda=$1545 nm.

  The observed pulses are identified as SR based on the following observations:
 (i) short pulse duration, 
 (ii) sech-squared pulse shape,
(iii) relationship between pulse duration and height, and 
(iv) relationship between pulse area and height.
  Below we elaborate these points in more detail.
Figure \ref{fig:SR} (c) is a typical example of a single observed pulse (the red crosses). 
  The FWHM (Full-Width-Half-Maximum) pulse duration of such pulses are observed to fluctuate  
  with the mean duration of 20 ns and the standard deviation of 5 ns, respectively.
  They are given by the Gaussian fit to the histogram of the duration.
  The observed pulse duration is $10^{6}$ times shorter than 
  the lifetime of the higher state of this transition ($\sim$ 10 ms), 
  whereas it is consistent with an expected time scale of SR. 
  Actually, the characteristic time of SR is given in a simplified two-level model by  
  \cite{Benedict-textbook-1996}
  \begin{equation}
 	T_{\mathrm{R}} =8\pi/(3 \gamma N_{0} \lambda^2 L) , 
        \label{eq:Tr}
  \end{equation}
  where $N_{0}$ denotes a number density of Er$^{3+}$ ions related to SR, 
  $\lambda$ is a wavelength of SR (1545 nm), $L$ is a sample length (6 mm), and 
  $\gamma$ is a radiative decay rate. 
  An approximate value of $T_{\mathrm{R}}$ may be estimated as follows. 
  For $N_0(=N/V)$, we use the number of photons $N$
  in a single pulse divided by the excitation volume $V$,  
  where $N\simeq\mathcal{O}(10^{12})$ is 
  estimated from the measured signal pulse area with the efficiency of the InGaAs photodetector, 
  and the transmittance of the filters, and 
  $V$ is taken to be $L$ times the laser beam cross section ($\pi w_0^2$).
  The upper limit of $\gamma$ is given by 
  the lifetime of the higher state multiplied 
  by the radiative branching ratio to the lower states \cite{ErYSO-spectroscopy-2006}, 
  and is found to be 40 Hz. 
  Combining these factors, $T_{\mathrm{R}}$ is calculated to be longer than $\mathcal{O}$ (1 ns). 
  Considering a numerical factor needed to convert $T_{\mathrm{R}}$ to the pulse duration and 
  ambiguity in $\gamma$, 
  we conclude that our observed pulses are consistent with SR.

 For our experimental configuration, it is expected that the macroscopic polarization 
 (coherence) develops homogeneously over the target 
 ($L \ll cT_{\mathrm{R}}$), and thus 
 SR pulse shapes become sech-squared function \cite{pure-oscillatory-SF-1975}. 
 As shown by the blue solid line in Fig. \ref{fig:SR} (c), the observed pulse shape is indeed 
 well fit by this function. 
 It is also expected that the pulse duration is inversely proportional to the square root of the peak height 
 since the former is proportional to $T_{\mathrm{R}}\propto 1/N$ (Eq.\ref{eq:Tr}), and 
 the latter to $N^2$.  
 Similarly the pulse area ($\propto N$) is proportional to the square root of the peak height. 
 These relationships are examined by the data and found to hold approximately 
 (see Supplemental Material for more details). 


  The periodic behavior of the observed SR 
  can be interpreted with the following simple model.
  (i)  The excitation laser pumps up Er$^{3+}$ ions to the higher state of the SR transition.
  (ii) A part of them deexcites to the lower states via spontaneous emissions. 
  (iii) During the process, macroscopic polarization (coherence) also develops, until 
       it reaches the SR threshold.
  (iv) Then, the population of the higher state 
       (also the coherence) suddenly decreases, resulting in a sharp pulse. 
  (v) The population of the lower state of SR decays to the ground state immediately.
  (vi) By repeating these processes, the periodic SR pulses are generated.

\begin{figure}
\centering
\includegraphics[width=7.5cm]{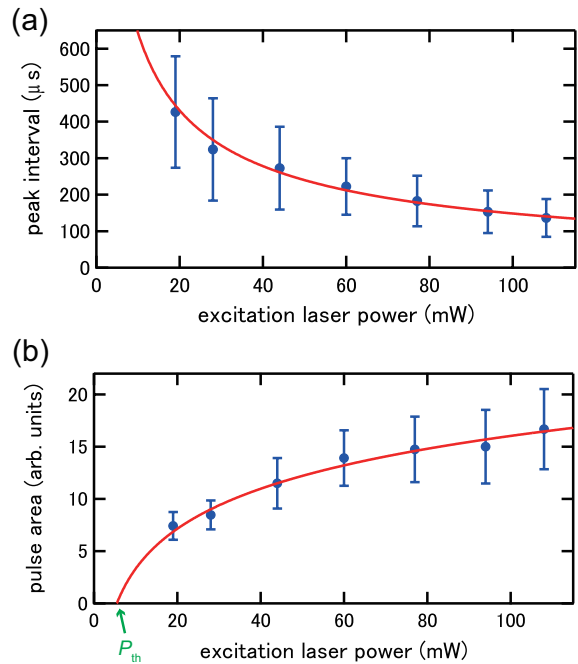}
\caption{(a) Peak interval at various excitation laser power.
The blue circles represent the averages.
The error bars indicate the standard deviations, not the statistical errors.
The red line is the fit of 
$(\mathrm{interval}) = A \times \ln (P/P_{\mathrm{th}})/(P-P_{\mathrm{th}})$ to the data.
(b) Pulse area at various excitation laser power.
The blue circles and the error bars indicate the averages and the standard deviations, respectively.
The red line is the fit of 
$(\mathrm{area}) = B \times \ln (P/P_{\mathrm{th}})$ to the data.
The threshold power $P_{\mathrm{th}}$ is common to both fits.}
\label{fig:period}
\end{figure}

To confirm the validity of our model, 
we investigated the dependence on the excitation laser power.
Figure \ref{fig:period} (a) and (b) show, respectively, the peak interval and pulse area  as a function of 
  the excitation laser power.
  All the observed pulses in 40 ms are used in this analysis.
  As the power increases, the period gets shorter and the area becomes larger.
The solid lines in Figs. \ref{fig:period} (a) and (b) are the results of fits 
  whose functional forms are derived from our model plus the assumption that 
  the SR threshold (thus $T_{\mathrm{R}}$) is independent of the laser power. 
  Since the coherence build-up is a competing process between pumping and deexcitation rates, 
  there should exist some threshold intensity $I_{\mathrm{th}}$ for the excitation laser. 
  The actual laser beam is not uniform spatially but has a radius dependence. 
  Thus an effective area in which the intensity exceeds $I_{\mathrm{th}}$ should increase as the 
  laser power. 
  Assuming a Gaussian intensity profile $I(r) 
  = I_{0} \exp(-2r^2/w_0^2)$, 
    we expect the effective volume 
$V_{\mathrm{eff}} = \pi r_{\mathrm{th}}^{2} L$ 
also increases as $\ln(I_0/I_{\mathrm{th}})$. 
Here $I_{0}=2P/\pi w_{0}^{2}$ is a peak intensity for the excitation laser power $P$ 
  and $r_{\mathrm{th}}$ is a threshold radius determined by $I (r_{\mathrm{th}}) = I_{\mathrm{th}}$.
  From the constancy of $T_{\mathrm{R}}$ or $N_0=N/V_{\mathrm{eff}}$, not $N/V$ here, we expect 
  $N$ (proportional to the pulse area) to increase as $\ln(I_0/I_{\mathrm{th}})$. 
  The peak interval, on the other hand, is inversely proportional to the intensity 
  averaged over the cross sectional area $\pi r_{\mathrm{th}}^2$ in our model: thus 
  we expect it is proportional to $\ln(I_0/I_{\mathrm{th}})/(I_0 - I_{\mathrm{th}})$. 
  Actual fits to the data, both peak interval and pulse area, are done simultaneously 
  using $\mathrm{(interval)}=A\times \ln(P/P_{\mathrm{th}})/(P-P_{\mathrm{th}})$ and  
  $\mathrm{(area)}=B\times \ln(P/P_{\mathrm{th}})$ 
  with $A,\;B$ and $P_{\mathrm{th}}$ as free parameters.
The threshold power $P_{\mathrm{th}}$ has a relationship of 
$I_{\mathrm{th}}=2P_{\mathrm{th}}/\pi w_{0}^{2}$ with $I_{\mathrm{th}}$.
The measured results are fit well by using our model.
The fitting results give $P_{\mathrm{th}} = 5.5 \pm 0.6$ mW.
This error is scaled so that the reduced $\chi^{2}$ value is unity. 
The value of $P_{\mathrm{th}}$ is consistent with the experimental fact 
that no periodic SR can be observed below 10 mW.
The above results of the excitation laser power dependence support the validity of our model.


\begin{figure}
\centering
\includegraphics[width=7.5cm]{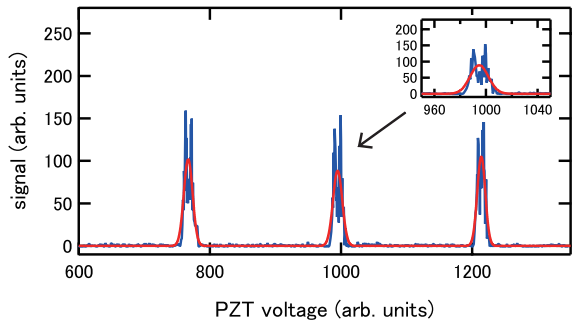}
\caption{Linewidth measurement of SR using a confocal cavity.
The cavity length is scanned by changing a PZT voltage.
The red line is the Gaussian fit.
The inset shows an enlargement of one of the peaks.}
\label{fig:linewidth}
\end{figure}

We now focus on the linewidth of SR. 
For this measurement, 
the SR pulses coming out of the crystal are coupled to an optical fiber, 
which is in turn sent to a confocal cavity (SA200-14; Thorlabs) 
whose Finesse and free spectral range (FSR) are $>200$ and 1.5 GHz, respectively.
While the cavity length is swept over 3 FSR slowly (at 0.013 Hz) by changing the PZT voltage, 
 the light intensity transmitted from the cavity is measured 
 by the same photodetector as in the SR experiment.
Figure \ref{fig:linewidth} is an example of the transmitted signal 
vs the cavity length.
The red line is the Gaussian fit, 
 which gives the FWHM linewidth of 100 MHz when averaged over three peaks. 
 Several comments are in order here. 
 First, the actual linewidth may be much narrower than the quoted value because 
 it contains shot-by-shot frequency fluctuations of the SR pulse. 
 In fact, the measured linewidth is broader than its Fourier limit, $\sim$20 MHz, which is
 estimated from the observed pulse width of $\sim$20 ns.
 Second, the 
measured width turns out to be much narrower 
than the typical inhomogeneous broadening in the crystal ($\sim$1 GHz).
 This result indicates that 
only the Er$^{3+}$ ions in a specific environment (with a narrow frequency range) are excited.
If Er$^{3+}$ ions were excited in a wider range of the inhomogeneous broadenings, 
we expect that each SR pulse would be generated independently, and thus randomly in time. 
We believe that the periodic nature of observed SR pulses is deeply related to this feature.


As a point of caution, 
the period changes depending on the excited region of the crystal due to inhomogeneity.
It is confirmed by changing the alignment of the excitation laser.
However, the periodic behavior can be reproduced if the alignment is adjusted properly.
Even if the period itself changes, 
the various properties of the periodic pulses can be discussed in the same way as above 
at other regions of the crystal.

In summary, we observed a phenomenon which can be characterized as a periodic SR  in an Er:YSO crystal.
The periodic time structure can be understood 
by our simple model, 
which presumes that the periodic SR is generated in a cyclic process of 
 continuous population inversion and 
 a sudden burst of SR emissions initiated at the SR threshold.
The validity of our model is supported, at least partially, by 
 the results of the power dependence measurements.
The model is admittedly qualitative: if more 
quantitative understanding 
is succeeded in, 
the control of the temporal behavior of the SR can be realized in the future. 
In addition to these features above, we revealed that, from the linewidth measurement,  
a narrow transition frequency range 
is selected for SR pulses within an inhomogeneous broadening.
We believe that this is one of the key features to realize the periodic time structure.
The periodic SR we observed also 
presents a potential as a light source, where narrow linewidth optical pulses are generated in 
a simple system without injection seeding.
Coherent phenomena in a solid have possible applications 
towards quantum information processing 
\cite{quantum-memory-2010,RE-quantum-memory-2015,RE-quantum-memory-2018} 
and fundamental physics \cite{axion-2018,RANP-2019,MagRENP-2021}.
Understanding and controlling coherent phenomena will play an important role 
in these research fields.

\begin{acknowledgments}
We thank C. Braggio, F. Choissi, G. Carugno, and K. An for helpful discussions.
This work was supported by JSPS KAKENHI (Grant Numbers JP20H00161 and JP21H01112).
\end{acknowledgments}


\providecommand{\noopsort}[1]{}\providecommand{\singleletter}[1]{#1}%

\end{document}



\title{Supplemental Material for: \\
Periodic super-radiance in Er:YSO crystal}


\author{Hideaki~Hara}
%
\email{hhara@okayama-u.ac.jp}
\affiliation{Research Institute for Interdisciplinary Science, Okayama University, Okayama 700-8530, Japan}
\author{Junseok~Han}
\affiliation{Research Institute for Interdisciplinary Science, Okayama University, Okayama 700-8530, Japan}
\affiliation{Department of Physics and Astronomy \& Institute of Applied Physics, Seoul National University, Seoul 08826, Korea}
\author{Yasutaka~Imai}
\affiliation{Research Institute for Interdisciplinary Science, Okayama University, Okayama 700-8530, Japan}
\author{Noboru~Sasao}
\affiliation{Research Institute for Interdisciplinary Science, Okayama University, Okayama 700-8530, Japan}
\author{Akihiro~Yoshimi}
\affiliation{Research Institute for Interdisciplinary Science, Okayama University, Okayama 700-8530, Japan}
\author{Koji~Yoshimura}
\affiliation{Research Institute for Interdisciplinary Science, Okayama University, Okayama 700-8530, Japan}
\author{Motohiko~Yoshimura}
\affiliation{Research Institute for Interdisciplinary Science, Okayama University, Okayama 700-8530, Japan}
\author{Yuki~Miyamoto}
\email{miyamo-y@cc.okayama-u.ac.jp}
\affiliation{Research Institute for Interdisciplinary Science, Okayama University, Okayama 700-8530, Japan}



\date{\today}



\maketitle



\section{Identification of the observed pulses as super-radiance}
\label{sec:prop}

In the main text in this paper, we identified the observed optical pulses 
as super-radiance (SR) 
based on the following observations: 
(i) short pulse duration, 
(ii) sech-squared pulse shape, 
(iii) relationship between pulse duration and height, and 
(iv) relationship between pulse area and height.
Here we discuss in more detail.

The FWHM duration of the observed pulses is $\mathcal{O} (10\mathrm{ns})$.
The emission process is accelerated by a factor of $10^{6}$ 
compared to the lifetime of the higher state of the transition ($\sim$ 10 ms).
The characteristic time of SR $T_{\mathrm{R}}$ is longer than $\mathcal{O} (1 \mathrm{ns})$.
Considering a numerical factor needed to convert $T_{\mathrm{R}}$ to the pulse duration 
and ambiguity in $\gamma$ (radiative decay rate), the observed pulse duration is consistent with SR.

The observed pulse shown in Fig. 2 (c) can be fit well by the sech-squared function 
$A \times \mathrm{sech}^{2} ((t-t_{0})/B)$, 
with $A$, $B$, and $t_{0}$ as the free parameters.
If the macroscopic polarization (coherence) develops homogeneously over the target ($L<<cT_{\mathrm{R}}$), 
the generated optical pulses go out of the sample so fast 
that they cannot be reabsorbed by target ions.
On the other hand, the target length is required to be much longer 
than a threshold length $L_{\mathrm{T}}$, 
which is derived from the requirement that the macroscopic dipole has to develop 
within a decoherence time $T_{2}$ \cite{pure-oscillatory-SF-1975}.
This requirement is satisfied because intense optical pulses are indeed observed.
The solution of the Maxwell-Bloch equations \cite{Benedict-textbook-1996} 
reduces to a sech-squared function without any ringings 
in the condition where $L_{\mathrm{T}}<<L<<cT_{\mathrm{R}}$.
This condition 
is satisfied in our system.

\begin{figure}
\includegraphics[width=8.5cm,keepaspectratio]{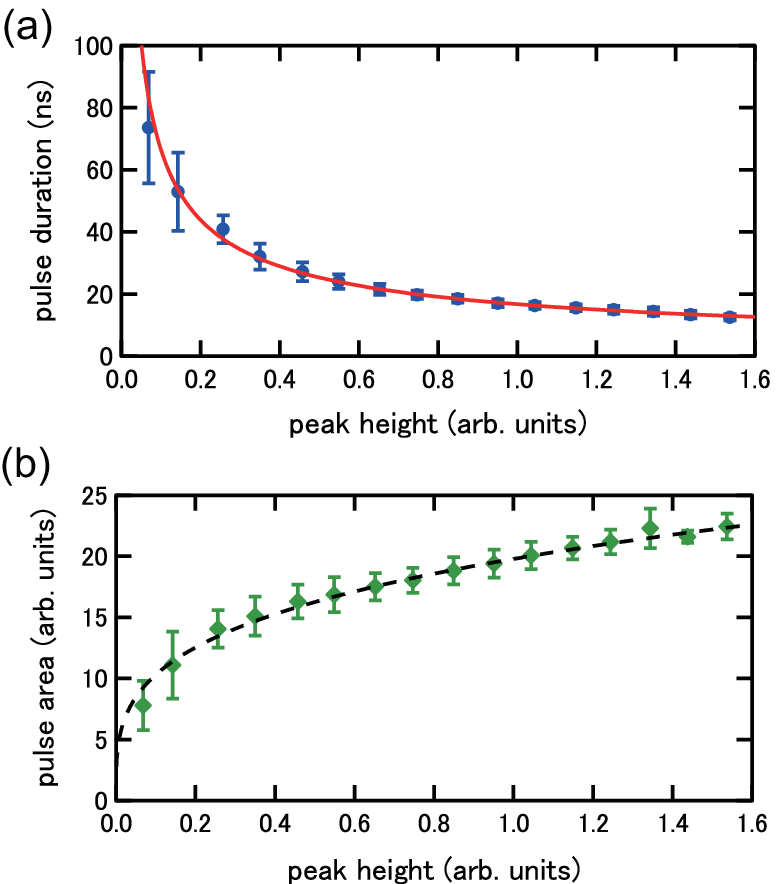}
\caption{(a) Relationship between pulse duration and peak height of each pulse.
The data points are taken only after $t=10$ ms.
The blue circles represent average of pulse duration in a certain peak height range.
The error bars indicate the standard deviations, not the statistical errors.
The red solid line is the power fit of $(\mathrm{duration}) = A \times (\mathrm{height})^{\alpha}$ to the data.
(b) Relationship between pulse area and peak height of each pulse.
The green diamonds and the error bars represent 
average of pulse area in a certain peak height range and the standard deviations, respectively.
The black dashed line is the power fit of $(\mathrm{area}) = A \times (\mathrm{height})^{\alpha}$ to the data.}
\label{fig:eachpulse}
\end{figure}

Fluctuations of the peak height, pulse duration, photon number are large 
due to the stochastic nature of SR.
Figure \ref{fig:eachpulse} (a) is the relationship between a peak height and a pulse duration.
The blue circles represent average of pulse duration in a certain peak height range.
The pulse duration increases with decrease in the peak height.
According to the spatially homogeneous solution of the Maxwell-Bloch equations, 
the pulse duration ($\propto 1/N$) 
is inversely proportional to the square root of the peak height ($\propto N^{2}$), 
where $N$ is the number of photons in a single pulse.
The red solid line is the power fit of $(\mathrm{duration})=A \times (\mathrm{height})^{\alpha}$ to the data, 
with $A$ and $\alpha$ as the free parameters.
The fitting result gives the index $\alpha = -0.60 \pm 0.01$, 
which is not significantly different from $-0.5$ of the spatially homogeneous solution.
Here the error is scaled so that the reduced $\chi^{2}$ value is unity.

Figure \ref{fig:eachpulse} (b) is the relationship between a peak height and a pulse area.
The pulse area is proportional to the total emitted photon number.
The green diamonds represent average of pulse area in a certain peak height range.
The pulse area increases together with the peak height.
According to the spatially homogeneous solution, 
the emitted photon number ($\propto N$) 
is proportional to the square root of the peak height ($\propto N^{2}$).
The black dashed line is the power fit of $(\mathrm{area}) = A \times (\mathrm{height})^{\alpha}$ to the data, 
with $A$ and $\alpha$ as the free parameters.
The fitting result gives the index $\alpha = 0.28 \pm 0.01$, 
where the error is scaled so that the reduced $\chi^{2}$ value is unity. 
The value of $\alpha$ is somewhat deviated from 
0.5 of the spatially homogeneous solution, 
but the expected relationship between a peak height and a pulse area is considered to hold approximately.
The result also supports the validity of the spatially homogeneous solution 
in that it deviates further from the value for the spontaneous emission ($\alpha = 1$, as mentioned later).

Because of the realistic experimental condition, 
it is possible that the results deviate from the simplified model, 
where the deexcitation process starts from a complete population inversion 
and the number of the excited ions are constant.
The source of the deviations from the spatially homogeneous solution is not clear currently.
On the other hand, in Ref. \cite{ErYSO-Padova-2020}, 
the authors explained the relationship between the peak photon emission rate and the observed photon number 
with assumption that 
the observed photon number is greater than the number of Er$^{3+}$ ions participating in the macroscopic state 
due to the single pass gain through the inverted medium.
Our result in Fig. \ref{fig:eachpulse} can also be fit by the functions derived in their model.

The pulses we observed are consistent with the spatially homogeneous solution of the Maxwell-Bloch equations.
Here we exclude another possibility of accelerating the deexcitation process, 
that is amplified spontaneous emission (ASE).
Even if coherence is not sufficiently developed, 
spontaneous emission is amplified as the radiation propagates through an inverted medium.
Firstly, the observed pulse shape is different from that of ASE.
The duration of the observed pulses is 20 ns on average, which is 
close to the order of the shorter limit of $T_{\mathrm{R}}$ $\mathcal{O} (1 \mathrm{ns})$.
The ASE pulse duration is expected to be much longer than $T_{\mathrm{R}}$.
The standard deviation of the observed pulse duration is 5 ns, 
which is much smaller than the variation of the ASE pulse duration.
In ASE, 
CW components comparable to the pulse components 
are also present \cite{SRtoASEtransition-1987}.
In our experiment, however, the CW component of the photons emitted in the forward direction 
is negligibly small with respect to the pulse component.
Secondly, the relationship between the pulse area and height is different from that of ASE.
In the case of spontaneous emission, the peak photon emission rate is proportional to the emitted photon number.
This case corresponds to $\alpha=1$ in $(\mathrm{area}) = A \times (\mathrm{height})^{\alpha}$.
In the case of ASE, the index $\alpha$ is expected to be between 0.5 (SR) and 1 (spontaneous emission).
Our result ($\alpha = 0.28 \pm 0.01$) 
is different from this range and is close to the limit of SR.
These results suggest that the observed pulses are not ASE.
From the above discussions, 
it is possible to identify that the pulses we observed are SR.

\begin{acknowledgments}
\end{acknowledgments}


\providecommand{\noopsort}[1]{}\providecommand{\singleletter}[1]{#1}%